\documentclass{article}
\usepackage[centertags]{amsmath}
\usepackage{amsfonts}
\usepackage{amssymb}
\usepackage{amsthm}
\usepackage{newlfont}
\usepackage{graphicx}
\usepackage{color}

\begin{document}

\renewcommand{\r}{\mathbf{r}}
\newcommand{\cn}{\mathop{\mathrm{cn}}}
\newcommand{\sn}{\mathop{\mathrm{sn}}}
\newcommand{\cs}{\mathop{\mathrm{cs}}}
\newcommand{\dn}{\mathop{\mathrm{dn}}}
\renewcommand{\sc}{\mathop{\mathrm{sc}}}
\newcommand{\am}{\mathop{\mathrm{am}}}

\title{Finite-Thickness and Charge Relaxation in Double-Layer Interactions}
\date{}
\author{Aldemar Torres and Ren\'e van Roij \\
Institute for Theoretical Physics, Utrecht University \\
Leuvenlaan 4, 3584 CE Utrecht, The Netherlands. \and Gabriel
T\'ellez\\Departamento de F\'\i sica, Universidad de Los Andes\\
A.~A.~4976 Bogot\'a, Colombia. }

\maketitle

\begin{abstract}
We extend the classical Gouy-Chapman model of two planar parallel
interacting double-layers, which is used as a first approximation
to describe the force between colloidal particles, by considering
the finite-thickness of the colloids. The formation of two
additional double layers due to this finite thickness, modifies
the interaction force compared to the Gouy-Chapman case, in which
the colloids are semi-infinite objects. In this paper we calculate
this interaction force and some other size-dependent properties
using a mean field level of description, based on the
Poisson-Boltzmann (PB) equation. We show that in the case of
finite-size colloids, this equation can be set in a closed form
depending on the geometrical parameters and on their surface
charge.  The corresponding linear (Debye-H\"uckel) theory and the
well-known results for semi-infinite colloids are recovered from
this formal solution after taking appropriate limits. We use a
density functional corresponding to the PB level of description to
show how in the case when the total colloidal charge is fixed, it
redistribute itself on their surfaces to minimize the energy of
the system depending on the afore mentioned parameters. We study
how this charge relaxation affects the colloidal interactions.
\end{abstract}

\section{Introduction}
Electrostatic forces play a remarkable role in the stabilization
and phase behavior of colloidal suspensions. Thus, it is not
surprising that understanding the interaction among charged
colloidal particles across an intervening electrolyte solution has
been an important goal for colloid science since the pioneering
work by Verwey and Overbeek~\cite{Verwey and Overbeek}.
Poisson-Boltzmann theory (PB) allows a mean field calculation of
these interaction at a nonlinear level. The corresponding
linearized description, Debye-H\"uckel theory (DH), is exact in
the asymptotic limit of infinite dilutions and low-charged
colloids, when the electrostatic interactions are weak compared to
the thermal energy. DH theory fails to describe quantitatively the
thermodynamic properties of realistic colloidal suspensions, and
nonlinear PB equation must be employed, often at the expense of
numerical computations.

\begin{figure}
\includegraphics[width=65mm]{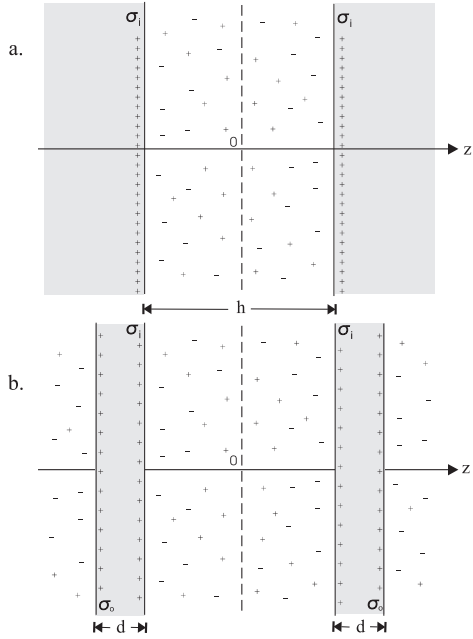}
\caption{ \label{fig:double-layers} a) Two parallel double-layers.
b) Four double-layers as a model for colloidal interactions. Here
the finite size of the colloids is taken into account.}
\end{figure}

Nonetheless, the PB equation admits analytical solution in some
non-trivial cases. An important example is that in which two parallel
infinite double-layers interact. In this case the geometry is dictated
by two planar objects (colloids) which are infinite in the directions
parallel to their surface and semi-infinite in one direction
perpendicular to their surface, and carrying a surface charge
$\sigma$. The region between the semi-infinite objects is filled with
symmetric electrolyte as illustrated in
figure~\ref{fig:double-layers}a. Thus, the force between the colloids
is mediated by the diffuse layers surrounding their surface. PB theory
allows the calculation of this force as well as the free energy of
interaction, through the determination of the mean local electrostatic
potential in the region filled with electrolyte \cite{Verwey and
Overbeek}. The planar geometry is relevant for colloidal spheres in
the case that the Debye length of the electrolyte is much smaller than
the radius of the spheres (such that the curvature effects are
unimportant). It is also relevant for understanding of colloidal clay
platelets and their interactions. An important aspect of these clay
platelets is their finite thickness $d$ which can easily be of the
same order as the Debye length and which could therefore affect the
plate-plate interactions.

The geometry of intend in this paper is depicted in figure 1b,
showing the finite thickness of the platelets and hence two inner
surfaces (with charge density $\sigma_i$) and two outer surfaces
(with charge density $\sigma_o$). Since the plates are completely
immersed in electrolyte, double-layers form at both sides of the
colloidal particles. The presence of charge on the outer surfaces
induces the redistribution of ions in the electrolyte, and diffuse
layers of charge appear in the vicinity of each region in contact
with the colloids. In other words, the finite-size of the
colloidal particles allows the formation and interaction of four
double-layers instead of two. It is clear that in this case, the
force between the colloidal particles depends not only on their
relative separation $h$ but also on their thickness $d$.

In the traditional model of two interacting double layers~\cite{Verwey
and Overbeek}, the initial studies supposed a fixed surface charge
density or a fixed potential at the surfaces.  But there have been
several studies in which the surface charge of the colloidal planes is
not fixed, but a charge regulation mechanism can occur~\cite{Ninham,
Behrens-Borkovec}, due to an ion adsorption-desorption of the
colloidal surface groups. When the finite-thickness of the colloids is
taken into account, as in our model (see
figure~\ref{fig:double-layers}b), there is another kind of charge
regulation that can take place. The total charge of each colloid can
be fixed but the amounts of charge on the inner and outer surfaces can
vary in order to minimize the thermodynamic grand potential of the
system. We call this mechanism ``charge relaxation''. This charge
relaxation also affects the force between the colloids. In this
contribution, we study quantitatively these finite-size effects in the
case of two equally charged colloids immersed in a symmetrical
electrolyte, in the framework of PB theory. The present nonlinear
study already accounts for charge renormalization effects
\cite{alexander,trizac-PRL,bocquet-JCP,trizac}.

In section~\ref{sec:model} we specify the model and introduce a
suitable density functional. In section~\ref{sec:PB}, the
nonlinear PB equation is solved. We show that for the geometry of
figure~\ref{fig:double-layers}b, the solution of the PB equation
can be set in a convenient analytical closed form, which depends
only on the geometrical parameters $h$ and $d$ and on the total
surface charge density $\sigma$ of the colloids. The linear DH
potential is obtained from the formal analytical solution by
taking the appropriate limits. In section~\ref{sec:charge-relax},
we minimize the density functional to calculate the fraction of
charge on the inner and outer surfaces, as a function of the afore
mentioned parameters, when charge relaxation is allowed. Results
are discussed in section~\ref{sec:results} and conclusions are
given in section~\ref{sec:conclusion}.

\section{The model}
\label{sec:model}

We consider four infinite parallel planes located at $z_1=-h/2-d$,
$z_2=-h/2$, $z_3=h/2$ and $z_4=h/2+d$, with surface charge
densities $\sigma_1$, $\sigma_2$, $\sigma_3$, $\sigma_4$
respectively.  The regions $z_1<z<z_2$ correspond to the interior
of the left colloid, and $z_3<z<z_4$ to the interior of the right
colloid. The regions outside the colloids are filled with an
electrolyte solution of dielectric constant $\epsilon$, at
temperature $T$. Inside the colloids there is no electrolyte
present but there is a medium with the same dielectric constant as
the electrolyte solution to avoid electrostatic image effects. The
system is assumed to be in contact with a salt reservoir of salt
density $\rho_s$ such that the electrolyte has screening length
$l_D=\kappa^{-1}=(8\pi \lambda_B \rho_s)^{-1/2}$, with
$\lambda_B=e^2/(k_B T \varepsilon)$ the Bjerrum length, where $e$
is the elementary charge, $k_B$ Boltzmann constant, and
$\varepsilon$ the electric permittivity of the solvent. In what
follows, all lengths will be conveniently given in units of the
Debye length, namely $D=\kappa{d}, H=\kappa{h}$ and $x=\kappa{z}$.
For simplicity we consider the case when
$\sigma_1=\sigma_4=\sigma_o$ and $\sigma_2=\sigma_3=\sigma_i$ as
illustrated in figure~\ref{fig:double-layers}b. Note that with
these conditions the system possess mirror symmetry with respect
to the plane at $x=0$ sketched with dashed lines.

The thermodynamic and statistical properties of the system are
characterized by the dimensionless grand potential density
functional $\omega[\rho]\equiv\kappa\beta\Omega[\rho]/(2A\rho_s)$
where $A$ is the area of the planes and $\Omega$ is the grand
potential density functional. $\omega[\rho]$ is given by
\begin{equation}
\label{omega}
 \omega[\rho]=\omega_{id}[\rho]+\frac{1}{2}\int{dx\rho(x)\phi(x)}+2f(\chi\log\chi+(1-\chi)\log(1-\chi)).
\end{equation}
The first term in the right hand side is the ideal gas entropic
contribution of the ions in the solution,
\begin{equation}
\omega_{id}[\rho]=\frac{1}{2\rho_s}\sum_{\alpha=\pm}\int{dx}\rho_\alpha(x)\left(\log
\left(\frac{\rho_\alpha(x)}{\rho_s}\right)-1 \right) ,
\end{equation}
where $\rho_{+}(x)$ and $\rho_{-}(x)$ are the number density
profiles of the positive and negative salt ions respectively. The
second term of (\ref{omega}) accounts for the electrostatic
energy, with $\phi(x)=e\psi(x)/k_BT$ the reduced electrostatic
potential ($\psi$ is the electrostatic potential), and the total
number charge in the system defined as
\begin{equation}
\rho(x)=\rho_+(x)-\rho_-(x)+\frac{\kappa}{e}\sum_{j=1}^{4}{\sigma_j}\delta(x-x_j).
\end{equation}
The last term in~(\ref{omega}) is of entropic origin and takes
into account the fact that the surface charge ions on the colloids
are indistinguishable. For later convenience we have introduced
$\chi=\sigma_i/\sigma$ which is the fraction of charge in the
inner surface with respect to the total charge density on a
colloid $\sigma=\sigma_i+\sigma_o$. Finally, the dimensionless
constant $f=4\pi\lambda_B\sigma/(e\kappa)$ is the ratio of the
Debye length $l_D=1/\kappa$ and the Gouy-Chapman length
$\Lambda=e/(4\pi\sigma\lambda_B)$. The later is the decay length
of the counterions distribution around an infinite charged plane
with bare surface charge density $\sigma$ in the salt free limit
$\rho_s\rightarrow0$ . Note that $f$ is an estimate of the
electrostatic energy compared to the characteristic energy of
thermal fluctuations in the system.

Minimization of the density functional (\ref{omega}) with respect
to the ion density profiles $\rho_{\pm}(x)$ gives the equilibrium
density profiles in terms of the average local potential. In the
mean field approach the reduced local electric potential $\phi(x)$
is the solution of the nonlinear Poisson--Boltzmann equation
\begin{equation}
\label{PB} \frac{d^2\phi}{dx^2}=\sinh \phi \qquad \text{in the
outside}
\end{equation}
\begin{equation}
\frac{d^2\phi}{dx^2}=0 \qquad \text{in the inside}
\end{equation}
with the boundary conditions that $\phi$ is a continuous function
of $x$ and
\begin{subequations}
\label{eq:BC}
\begin{equation}
\textstyle
\phi'(\frac{H}{2}^-)-\phi'(\frac{H}{2}^+)=f_i
\end{equation}
\begin{equation}
\textstyle
\phi'((\frac{H}{2}+D)^-)-\phi'((\frac{H}{2}+D)^+)=f_o
\end{equation}
\begin{equation}
  \lim_{|x|\to\infty}\phi'(x)=0
\end{equation}
\end{subequations}
with $f_i=4\pi l_B\sigma_i/(e\kappa)$, $f_o=4\pi
l_B\sigma_o/(e\kappa)$ as explained before.

The functional (\ref{omega}) evaluated for the equilibrium
profiles gives the equilibrium dimensionless grand potential.
Further minimization of the later quantity with respect to the
parameter $\chi=\sigma_i/\sigma$ allows to find the fraction of
charge on each surface, when charge relaxation is allowed. This
will be discussed in section~\ref{sec:charge-relax}. The DH level
of description can be obtained by expanding (\ref{omega}) to
second order in the density and linearizing (\ref{PB}) for
$|\phi|\ll1$. It can also be obtained from the nonlinear solution
as will be discussed next.

\section{Solution of Poisson-Boltzmann equation}
\label{sec:PB}

\subsection{The Nonlinear PB Equation}

Due to the mirror symmetry it is sufficient to consider $x\geq0$
since the potential is an even function of $x$:
$\phi(x)=\phi(-x)$. Also, without loss of generality, we consider
positively charged colloids i.e. $f_i>0$. The solution of the PB
equation in the region $-H/2<x<H/2$ with the boundary conditions
$\phi(0)=\phi_0$ and $\phi'(0)=0$ is well-known. There are several
different but equivalent expressions for this
solution~\cite{Verwey and Overbeek, Behrens-Borkovec,
McCormack-PB-solutions, Tamashiro-Schiessel}. After the usual
first integration of the PB equation and using the boundary
condition at $x=0$ we have
\begin{equation}
\label{phi prime}
  \phi'(x)^2=4\left(\sinh^2\frac{\phi(x)}{2}-\sinh^2\frac{\phi_0}{2}
  \right)
  \,.
\end{equation}
From this equation we see that
$\sinh^2\frac{\phi(x)}{2}>\sinh^2\frac{\phi_0}{2}$, thus
$|\phi(x)|>|\phi_0|$. We deduce that $\phi(x)$ does not change its
sign. Furthermore with the choice $f_i>0$, we have $\phi(x)\geq
\phi_0$, and then $\phi'(x)\geq0$.

Integrating once again with respect to $x$ and defining
$t=\sinh(\phi/2)$, the integral can be set in the form
\begin{equation}
x=\int_{\sinh(\phi_0/2)}^{\sinh(\phi(x)/2)}\frac{dt}{\sqrt{
    (t^2+1)(t^2-\sinh^2(\phi_0/2))}}
\,.
\end{equation}
One recognizes an elliptic integral of the first
kind~\cite{Gradshteyn}, defined as
\begin{equation}
  F(\theta,k)=\int_0^\theta \frac{d\alpha}{\sqrt{1-k^2\sin^2\alpha}}
  ,
\end{equation}
and hence we have
\begin{equation}
x=\frac{1}{\cosh(\phi_0/2)}
F\left(\arccos\frac{\sinh(\phi_0/2)}{\sinh(\phi(x)/2)},
\frac{1}{\cosh(\phi_0/2)}\right)
\,.
\end{equation}
An inversion yields
\begin{equation}
  \sinh\frac{\phi(x)}{2}=\frac{\sinh(\phi_0/2)}{\cn(x\cosh(\phi_0/2),k)}
  ,
\end{equation}
where $\cn$ is the Jacobian cosine amplitude. The modulus
$k=1/\cosh(\phi_0/2)$ of the elliptic functions will not be
explicitly indicated in the following to lighten the notation. To
take into account the finite-size of the colloids, we match this
expression with the solutions in the regions $H/2<x<H/2+D$ and
$H/2+D<x$. We get
\begin{equation}
  \label{eq:phi1}
  \phi(x)=\phi_1(x)=2 \sinh^{-1}\left[\frac{\sinh(\phi_0/2)}{
      \cn(x\cosh(\phi_0/2))}\right]\qquad\text{for\ }0\leq x\leq H/2
\end{equation}
\begin{equation}
  \label{eq:phi2}
  \phi(x)=\phi_2(x) = a \left(x-\frac{H}{2}\right)+\phi_1(H/2)
  \qquad\text{for\ }H/2 \leq x \leq H/2+D
\end{equation}
with the dimensionless electric field inside the colloids given by
\begin{equation}
  \label{eq:def-a}
  a=2 \sinh(\phi_0/2)\sc\left(\frac{H}{2}
    \cosh\frac{\phi_0}{2}\right)-f_i.
\end{equation}
For the outside region $x>H/2+D$ we find
\begin{equation}
  \label{eq:phi3}
  \phi(x)=\phi_3(x)=2\ln\frac{1+Ce^{-x}}{1-Ce^{-x}}
  \qquad \text{for\ }H/2+D\leq x,
\end{equation}
with
\begin{equation}
\label{C as Tanh}
  C=\tanh(\phi_2(
  \textstyle
  \frac{H}{2}+D)/4)e^{\frac{H}{2}+D},
\end{equation}
where $\sc(u,k)=\sn(u,k)/\cn(u,k)$ with $\sn(u,k)$ the Jacobian
sine amplitude~\cite{Gradshteyn, Whittaker-Watson}. For a fixed
$\phi_0$ equations (\ref{eq:phi1}), (\ref{eq:phi2}) and
(\ref{eq:phi3}) give the exact analytic form of the mean field
within the Poisson-Boltzmann theory. To completely specify the
spatial dependence of $\phi$, we still need to determine its value
at the origin $\phi_0$. For this we use the boundary conditions
(\ref{eq:BC}) to get
\begin{equation}
  \label{eq:for-phi0}
  \frac{f_o+f_i}{2}=
  \sinh\left[
    \frac{aD}{2}+\sinh^{-1}\left(\frac{\sinh(\phi_0/2)}{
      \cn(\frac{H}{2}\cosh\frac{\phi_0}{2})}\right)
    \right]
  +\sinh(\phi_0/2)
  \sc\left(\frac{H}{2}
    \cosh\frac{\phi_0}{2}\right)
\end{equation}
with $a$ given by eq.~(\ref{eq:def-a}).

For given values of the parameters $f_i$, $f_o$, $H$ and $D$,
equation ~(\ref{eq:for-phi0}) can be easily solved numerically for
$\phi_0$~\cite{Mathematica}. In combination with (\ref{eq:phi1}) to
(\ref{C as Tanh}) it gives the exact mean field as a function of
$f_i$, $f_0$, and the geometrical parameters $H$ and $D$.

An interesting feature of $\phi_1(x)$ as given in (\ref{eq:phi1})
is that if we consider the prolongation of the function $\phi_1$
in eq.~(\ref{eq:phi1}), for any value of $x$, from the general
properties of the elliptic function $\cn$, we notice that $\phi_1$
is a periodic function of $x$ with period $4K$, where
$K=F(\pi/2,k)$. For $x>0$, it has a first pole at the first
positive zero of $\cn$, at $x=K/(\cosh(\phi_0/2))$. Physically,
the potential should be bounded. This means that for a given
separation $H$, $\phi_0$ satisfies
\begin{equation}
\label{eq:Hmax} H<H_{\max}=\frac{2K}{\cosh(\phi_0/2)}.
\end{equation}
The case $H=H_{\max}$ corresponds to a situation where the surface
charge $\sigma_i\to+\infty$. Eq.~(\ref{eq:Hmax}) serves as a guide
to choose the initial value for $\phi_0$ in the numerical
resolution of~(\ref{eq:for-phi0}), or in a direct numerical
integration of Poisson--Boltzmann equation~(\ref{PB}).

\subsection{The Linear Regime: Debye--H\"uckel Theory}

We can recover the results from the linear Debye--H\"uckel theory by
taking the limit $\phi_0\ll 1$. In that limit the modulus of the
elliptic functions $k\to1$. Using $\cn(u,1)=1/\cosh u$, we have $
\phi_1(x)= \phi_0 \cosh x $ for $\phi_0\ll1$, which is the solution of
the linear problem in the region $0\leq x \leq H/2$.  Similarly, using
$\sn(u,1)=\tanh u$ we have
\begin{equation}
  \label{eq:a-linear1}
  a=\phi_0 \sinh(H/2) - f_i,
\end{equation}
and equation~(\ref{eq:for-phi0}) reduces to a linear equation
which can be easily solved for $\phi_0$. The result is
\begin{equation}
\label{phioDH}
  \phi_0=\frac{f_o+(1+D)f_i}{e^{H/2}+D\sinh (H/2)}.
\end{equation}
The last equation in combination with (\ref{eq:a-linear1}) and the
expression for $\phi_1(x)$, determines also the potential in the
region ${H/2}\leq{x}\leq{H/2+D}$, namely, $\phi
_{2}(x)=a(x-H/2)+\phi _{1}(H/2)$. Finally, linearizing
(\ref{eq:phi3}) and (\ref{C as Tanh}) we get $\phi
_{3}(x)=[aD+\phi_1(H/2)]e^{H/2+D}e^{-x}$. It is an easy exercise
to show that the same expressions can be obtained with the DH
approximation taken ab initio. For that, it is enough to take
functions of the form $\phi_1(x)=Ae^x+Be^{-x}$,
$\phi_2(x)=a(x-H/2)+\phi_1(H/2)$ and $\phi_3(x)=Ce^{H/2+D}e^{-x}$.
Using the boundary conditions and the continuity of the potential,
a system of coupled linear equations for the constants $A, B, C$
and $a$ is obtained. The solution coincides with the exact
nonlinear solution in the limit of small potentials.

\section{The Equilibrium Grand Potential}
\label{sec:charge-relax}
\begin{figure}
\label{omega-fig}
\includegraphics[width=65mm,angle=-90]{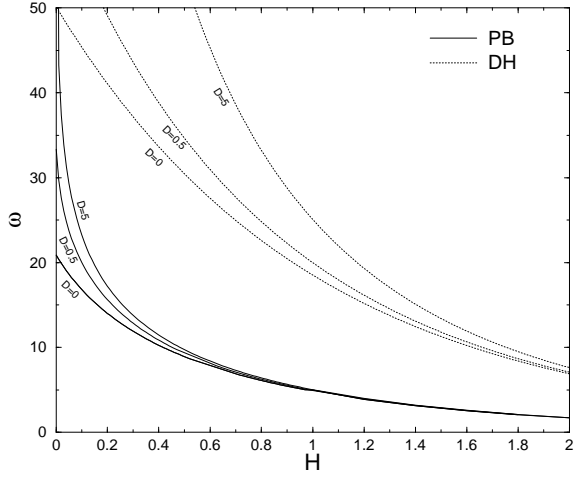}
\caption{Exact and DH based equilibrium grand potential as a
function of the separation distance H for two interacting
finite-size planes. In this plot $\chi=0.5$ and $f=10$.}
\end{figure}

As mentioned before, the total charge on the colloids is assumed
to be fixed, but eventually different fractions of charge can move
between the inner and outer surfaces. We took this into account in
the density functional  (\ref{omega}) by fixing the total charge
on the colloids instead of the surface charge on the inner and
outer surfaces separately, through the introduction of the charge
fraction $\chi$. The functional (\ref{omega}) evaluated at the
equilibrium density profile gives, for fixed $\chi$, the grand
potential of the system. Further minimization of the later
quantity with respect to $\chi$ allow us to find the fraction of
charge on each surface that minimizes the equilibrium grand
potential. The grand potential is obtained by inserting the
expressions ~(\ref{eq:phi1}), (\ref{eq:phi2}) and (\ref{eq:phi3})
into the density functional (\ref{omega}). The explicit
calculation involves well-known quadratures of the Jacobian
elliptic functions, in particular
\begin{equation}
  \int \frac{du}{\cn^2 (u,k)}=\frac{\sc (u,k) \dn
  (u,k)-E(\am(u,k),k)}{1-k^2} +u ,
\end{equation}
where $\dn(u,k)$ is the Jacobian delta amplitude, $E(x,k)$ the
elliptic integral of second kind and $\am(u,k)$ the Jacobian
amplitude~\cite{Gradshteyn, Whittaker-Watson}. After some
calculations we obtain
\begin{eqnarray}
  \label{eq:gp-res}
  \omega&=&
  2\phi(H/2)\sinh(\phi_0/2)\sc(H/2k)
  -8\cosh(\phi_0/2)\sc(H/2k)\dn(H/2k)
  \nonumber\\
  &&
  +8E(\am(H/2k),k)\cosh(\phi_0/2)
  -2H\sinh^2(\phi_0/2)
  \\&&
  \nonumber
  +2\sinh(\phi(\textstyle\frac{H}{2}+D)/2)\phi(\frac{H}{2}+D)
  -16\sinh^2(\phi(\frac{H}{2}+D)/4)
  \\&&
  \nonumber
  \textstyle
  +f[2\chi\ln\chi+2(1-\chi)\ln(1-\chi)
  +\chi\phi(H/2)+(1-\chi)\phi(\frac{H}{2}+D)].
  \nonumber
\end{eqnarray}
Notice that $k=1/\cosh(\phi_0/2)$, that $\phi(x)$ can be evaluated
from~(\ref{eq:phi1}), (\ref{eq:phi2}), (\ref{eq:phi3}), and that
$\phi_0$ is the solution of~(\ref{eq:for-phi0}) and thus depends
on $\chi$. For given values of $H$, $D$ and $f$ we determine  the
value of $\chi$ that minimizes the grand
potential~(\ref{eq:gp-res}) numerically~\cite{Mathematica}.

The DH level of description is obtained from the linear density
functional corresponding to (\ref{omega}). In this case the
dimensionless grand potential reads
\begin{equation}
\label{eq:omegaDH}
\omega=
  f[2\chi\ln\chi+2(1-\chi)\ln(1-\chi)
    \textstyle
  +\chi\phi(H/2)+(1-\chi)\phi(\frac{H}{2}+D)]
\end{equation}
where $\phi$ is given by the expressions derived in section 3.2.
Although equations (\ref{eq:gp-res}) and (\ref{eq:omegaDH}) are a
bit involved, the grand potential has the familiar repulsive form
as that for semi-infinite colloids. This is illustrated in figure
2, where the linear and nonlinear results are compared for the
case $f=10$ and $\chi=0.5$. Some consequences of the
size-dependence of the potential will be discussed in the next
section.

\section{Results}
\label{sec:results}

\subsection{Charge relaxation}
\begin{figure}
\includegraphics[width=65mm,angle=-90]{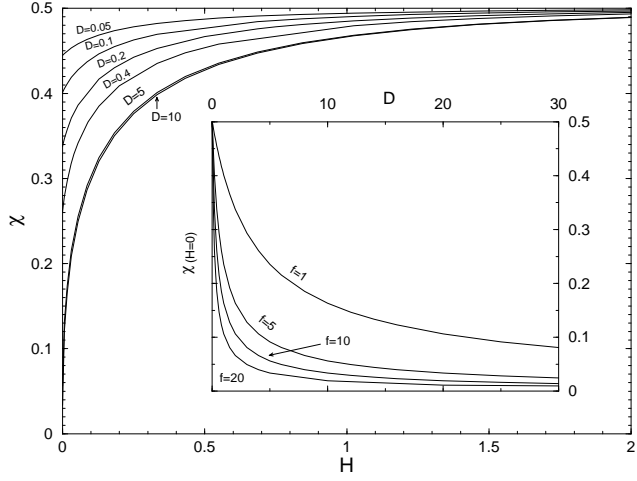}
\caption{ \label{fig:chi} Fraction of surface charge density
$\chi$ as a function of colloid separation for different colloidal
sizes for $f=10$. In the inset we see how $\chi({H}\rightarrow0)$
changes with varying $f$.}
\end{figure}

Let us first consider the aforementioned effect of charge
relaxation, in which charges can move between the colloidal
surfaces. We shall refer to this as the annealed case, in contrast
to the case in which equal amounts of charge are fixed
(``quenched'') on the inner and outer surfaces. Notice that in the
quenched case $\chi=\sigma_i/\sigma=0.5$. In figure~\ref{fig:chi},
the charge fraction $\chi$ is plotted as a function of the colloid
separation $H$ for different colloidal thicknesses. We observe
that for inter-colloidal separations of the order of the Debye
length, a fraction of charge moves to the outer surface. This
effect substantially increases with increasing values of the
parameter $f$, as we can conclude from the inset, which shows the
fraction of charge at $H\rightarrow0$. Notice that for increasing
$f$, the values of $\chi$ at contact tend faster to zero. Thus,
the effect of increasing $f$ is similar to what we observe in the
main plot when $D$ is increased, namely, a more pronounced
relaxation of charge toward the outer surface. Notice that this
effect already disappears at colloidal separations of the order of
two times the Debye length, at which a small difference with the
quenched case is observed. It also worth to mention the fact that
in the main plot, no substantial change is observed when the
thickness is increased from $D=5$ to $D=10$. In fact, for a given
value of $H\neq0$, when the thickness varies from $D=5$ to
$D=\infty$, the charge fraction $\chi$ does not change beyond 3\%
. For a colloidal thickness of a few Debye lengths, the distance
to the outer surface is already too large to affect the fraction
of charge that migrates as we vary the distance between the
colloids.

\subsection{Electrostatic potential}

\begin{figure}
\includegraphics[width=65mm,angle=-90]{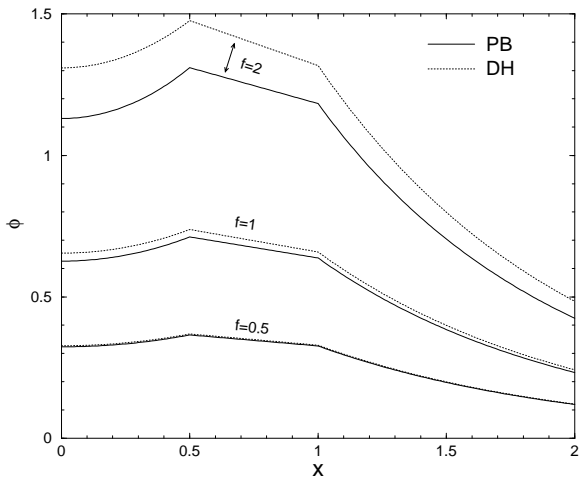}
\caption{ \label{fig:phi} Average local electrostatic potential
for $H=1$, $D=0.5$ and $\chi=0.5$ (quenched case). Different
values of the dimensionless parameter $f$ are plotted. Notice that
the linear DH results are asymptotically exact in the limit $f\ll
1$ but fail to quantitatively predict the local mean field for
$f\geq1$.}
\end{figure}

\begin{figure}
\includegraphics[width=65mm,angle=-90]{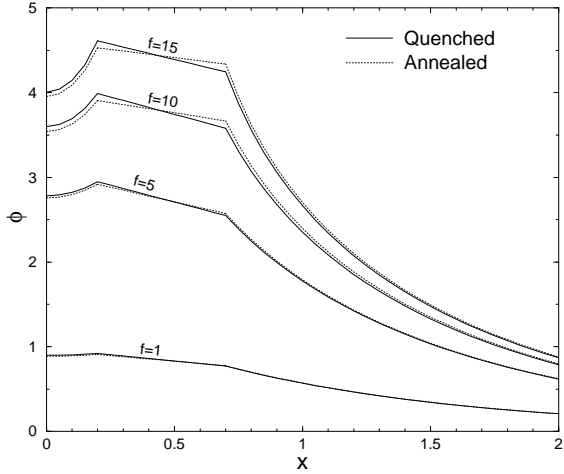}
\caption{ \label{fig:phiQA} Comparison of the local electrostatic
potential $\phi(x)$ for the quenched and annealed cases. The
geometric parameters are same as in figure~\ref{fig:phi}. Moderate
difference is only observed for $f\gg1$ in accordance with
figure~\ref{fig:chi}.}
\end{figure}

In figure~\ref{fig:phi} we plot the electrostatic potential
$\phi(x)$ for $x>0$ in the quenched case, obtained by equations
(\ref{eq:phi1}-\ref{eq:phi3}) using the methods described in
section~\ref{sec:PB}. The dotted line represents the corresponding
linear DH mean potential. Even though for $f\ll1$ both theories
coincide, DH theory fails to quantitatively predict the local
value of the mean field for $f\gtrsim 1$. For example, for
$D\rightarrow0$, (\ref{phioDH}) reduces to
$\phi_{0}=(f_{i}+f_{o})e^{-H/2}$, thus, the total charge density
$\sigma_{i}+\sigma_{o}$ is concentrated in a plane of
infinitesimal width and the potential at the plane of symmetry
decreases exponentially with respect to the plates separation. In
the limit $D\rightarrow\infty$ we have
$\phi_{0}=f_{i}/\sinh(H/2)$, this expression diverges for
$H\rightarrow0$ whereas eq.~(\ref{eq:for-phi0}) predicts a finite
value for this quantity. In figure~\ref{fig:phiQA}, we compare the
electrostatic field in the quenched case with that in the annealed
case. Notice that as charge moves from the inner to the outer
surface, the potential at the first decreases while it increases
at the later. In other words relaxation of charge reduces the
potential difference between the inner and outer surfaces. This
effect is stronger for larger $f$ in accordance with
figure~\ref{fig:chi}.

\subsection{Force between the colloids}

\begin{figure}
\includegraphics[width=65mm,angle=-90]{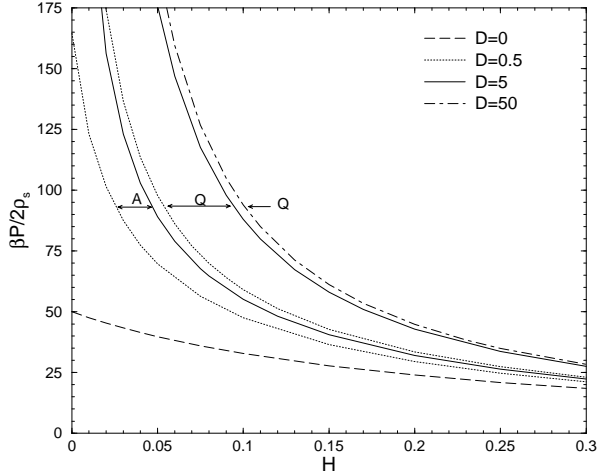}
\caption{ \label{fig:pressDsH} Dimensionless force per unit area
between the colloids for $f=10$. This quantity is always positive,
indicating repulsion between the colloids, and tends to zero for
as the separation between the colloids increases. The plots for
the quenched and annealed cases are labelled with Q and A
respectively. }
\end{figure}

The pressure is the force between the plates per unit area, and is
defined in terms of the grand potential through the relation $\beta
p/(2\rho_s)=-\partial\omega/\partial H$. Alternatively, taking into
account the fact that the electrostatic pressure of the system must
balance the hydrostatic pressure, it can be written in terms of the
mean field $\psi$ as~\cite{Verwey and Overbeek}
\begin{equation}
\label{pressure}
p= k_B T (n(x)-2\rho_s)-\frac{\varepsilon }{8\pi
}\left( \frac{d\psi }{dx}\right)^{2}+
\frac{\varepsilon }{8\pi }\left( \frac{d\psi
}{dx}\right) _{H\rightarrow \infty }^{2}
\end{equation}
where $n(x)=\rho_{+}(x)+\rho_{-}(x)$ is the total density of
microions at $x$. It is easy to show that within PB theory, this
quantity gives the same value for any position $x$. In particular
evaluating at $x=0$ we get
\begin{equation}
\label{p-phio} \beta p=2\rho _{s}(\cosh \phi _{0}-1),
\end{equation}
which implies that the force between the colloids is completely
determined by the value of the mean electrostatic potential at the
plane of symmetry. This feature allow us to find this quantity
directly from the numerical solution of equation
(\ref{eq:for-phi0}). In figure~\ref{fig:pressDsH} we plot the
dimensionless pressure as a function of the colloidal separation,
for different values of the thickness $D$ for the quenched and
annealed cases. Although the last two give obviously the same
curve for $D=0$, a substantial difference is observed for small,
nonzero, colloidal thicknesses. On the other hand, for $D=10$ and
larger, the inter-colloidal force turns out to be almost
independent of the thickness for the same reasons discussed before
in the context of charge relaxation. All the curves tend
asymptotically to zero in the limit of infinite colloidal
separation. In the DH limit $\beta p=2\rho _{s}(\cosh \phi
_{0}-1)\simeq \rho _{s}\phi _{0}^{2}$. Using (\ref{phioDH}) we
obtain $\beta p =\rho _{s}f_{i}^{2}/{\sinh ^{2}}\left(
\frac{H}{2}\right)$ for $D\rightarrow\infty$. Again, in the limit
$H\rightarrow0$, the linear theory predicts divergent behavior for
the pressure, while the full nonlinear theory gives a finite value
for this quantity.

\subsection{Average charge density between the colloids}

\begin{figure}
\includegraphics[width=65mm,angle=-90]{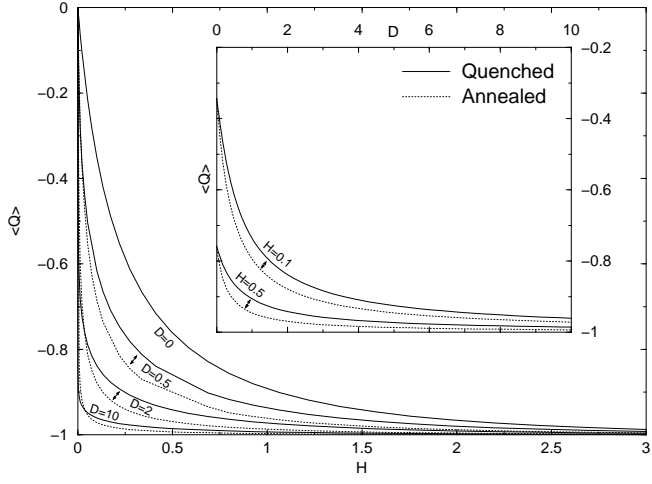}
\caption{ \label{fig:meanQ} Average charge density in the region
between the colloids for $f=10$. When $D\rightarrow\infty$ charge
of sign opposite to that on the colloidal surface is present in
the middle region to neutralize the charge on the colloids. When
$D$ is finite, part of that charge is squeezed out of the middle
region as the colloids are driven to a closer relative position.
The insert shows the squeezing of charge when we decrease the
thickness for fixed separation $H$.}
\end{figure}

Using Poisson's equation $\phi''(x)=-\rho(x)/(2\rho_s)$, we
compute the ratio $\langle Q\rangle$ of the average ion charge
density between the two colloids, compared to the charge density
on the inner faces of the colloids:
\begin{eqnarray}
\left\langle Q\right\rangle =\frac{e}{\kappa\sigma
_{i}}\int_{0}^{H/2}\rho(x)dx= \frac{1}{f_{i}}\left( \phi ^{\prime
}(0)-\phi ^{\prime }(\textstyle\frac{H}{2}^{-})\right)
=-\frac{\phi'(\frac{H}{2}^{-})}{f_i} .
\end{eqnarray}
This quantity is shown as a function of the inter-colloidal
distance for different values of $D$ in figure~\ref{fig:meanQ}.
When the distance between the finite-width colloids is shortened,
charge can be squeezed out of the region between the colloids.
This is not possible for semi-infinite colloids, since in that
case there is no place for the charge to be squeezed out, such
that $\langle Q\rangle=-1$ for all $H$ as shown in
figure~\ref{fig:meanQ}. The inset shows the squeezing of charge
when we decrease the thickness for fixed separation $H$. For $D=0$
the quenched and annealed cases coincide. One expects that Q is
closer to $-1$ in the annealed case than in the quenched case,
since the surface charge can follow the squeezed-out ionic charge
by migration to the outer surface. This is indeed shown by our
numerical results as shown in figure 7. However, for very small
separations, $H\approx10^{-2}$ this effect is reversed, as can be
seen in figure 8, where $\langle Q\rangle$ is closer to $-1$ in
the quenched case. Note that this effect occurs for rather small
values of the separation distance. In a model that accounts for
the finite-size of the ions this effect is presumably absent.

\begin{figure}
\includegraphics[width=65mm,angle=-90]{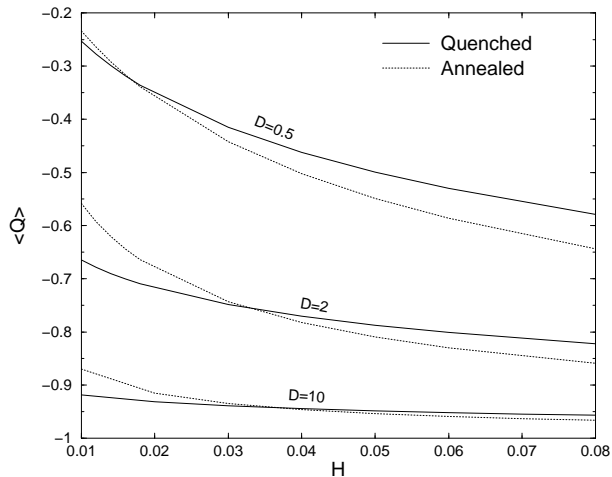}
\caption{ \label{fig:zoomQ} Average charge density (see text) in
the region between the colloids for $f=10$. }
\end{figure}

\section{Conclusion}
\label{sec:conclusion}

The system consisting of two interacting double-layers, used as a
model of colloidal interactions when the curvature of the
macromolecules is negligible has been extended to consider
finite-width colloids, by the inclusion of two additional
interacting double layers. The force that drives the colloidal
particles apart is modified by their finite-size. The nonlinear PB
equation was solved considering this finite-size to find
quantitatively the dependence of the mean field with respect to
the geometrical parameters. We considered two relevant cases: the
quenched case where the surface charge density on each surface of
the colloids is the same and the annealed case, where the total
charge of each colloid is fixed, but not the charge on each
surface. In the latter case a charge ``relaxation'' effect is
possible where the charges in each surface rearrange themselves in
order to minimize the grand potential of the system. We found that
this relaxation effect modifies also the mean electrostatic
potential, and as a consequence the force between the colloids. We
consider these results as a stepping stone towards the study of
more complicated charge-relaxation and charge-regularization
mechanisms, and to more complicated geometries, such as spheres
treated within the Derjaguin approximation.

\section{Acknowledgments}
This work is part of the research program of the "Stichting voor
Fundamenteel Onderzoek der Materie (FOM)," which is financially
supported by the "Nederlandse Organisatie voor Wetenschappelijk
Onderzoek (NWO)." G.~T.~acknowledges partial financial support
from COLCIENCIAS project 1204-05-13625.

%%%%%%%%%%%%%%%


\begin{thebibliography}{99}


\bibitem{Verwey and Overbeek}
E.~J.~W.~Verwey and J.~Th.~G.~Overbeek, \textit{Theory of the stability
of lyophobic colloids} (Elsevier, Amsterdam, 1948).

\bibitem{Ninham}
B.~W.~Ninham and V.~A.~Parsegian,
J.~Theor.~Biol.~\textbf{31}, 405 (1971).

\bibitem{Behrens-Borkovec}
S.~H.~Behrens and M.~Borkovec, Phys.~Rev.~E \textbf{60}, 7040 (1999).

\bibitem{alexander}
S. Alexander, P. M. Chaikin, P. Grant, G. Morales, P. Pincus, and
D. Hone, \textit{J. Chem. Phys.}, \textbf{80}, 5776 (1984).

\bibitem{trizac-PRL}
 E.~Trizac, L.~Bocquet and M.~Aubouy, 
 \textit{Phys.~Rev.~Lett.~}\textbf{89}, 248301 (2002).
 
\bibitem{bocquet-JCP}
 L.~Bocquet, E.~Trizac, M.~Aubouy,
 \textit{J.~Chem.~Phys.~}\textbf{117}, 8138 (2002).

\bibitem{trizac}
E. Trizac and Y. Levin, \textit{Phys. Rev. E}, \textbf{69}, 031403
(2004).

\bibitem{McCormack-PB-solutions}
D.~McCormack, S.L.~Carnie, and D.~Y.~C.~Chan,
J.~Colloid Interface Sci.~\textbf{169}, 177 (1995).

\bibitem{Tamashiro-Schiessel} M.~N.~Tamashiro and H.~Schiessel,
Phys.~Rev.~E \textbf{68}, 066106 (2003).

% No citado
%
%\bibitem{Barrat-Hansen}
%J. L. Barrat and J. P. Hansen, \textit{Basic Concepts for Simple
%and Complex Fluids} (Cambridge 2003).

\bibitem{Gradshteyn}
I.~S.~Gradshteyn, I.~M.~Ryzhik, \textit{Table of Integrals, Series,
and Products} (Academic Press 1994).

\bibitem{Mathematica} All the numerical resolutions (root finding and
minimization) involving the Jacobian elliptic functions were performed
using \textsc{Mathematica} version 5.

\bibitem{Whittaker-Watson} E.~T.~Whittaker and G.~N.~Watson, \textit{A
course of modern analysis}, (Cambridge, 1927)

\end{thebibliography}
\end{document}